\documentclass[twocolumn, prl]{revtex4-1}
\usepackage{graphicx}
\usepackage{epsfig}
\begin{document}
\title{Alternative Route to Strong Interaction:
 {\em Narrow} Feshbach Resonance}
\author{Tin-Lun Ho$^{\dagger}$, Xiaoling Cui$^{\dagger\ast}$, Weiran Li$^{\dagger}$ }
\affiliation{$^{\dagger}$Department of Physics, The Ohio State University, Columbus, OH 43210, USA\\
$^{\ast}$ Institute for Advanced Study, Tsinghua
University, Beijing 100084, China}
\date{\today}

\begin{abstract}
We show that a narrow resonance produces strong interaction effects far beyond its width on the  side of the resonance where the bound state has not been formed. This is due to a resonance structure of its phase shift, which shifts the phase of a large number of scattering states by $\pi$ before the bound state emerges. As a result, the magnitude of the interaction energy when approaching the resonance on the ``upper" and ``lower" branch from different side of the resonance is highly asymmetric, unlike their counter part in wide resonances. Measurements of these effects are experimentally feasible.
 \end{abstract}

\maketitle

Strongly interacting Fermi gases are known for their universal properties and their  ``high" superfluid  transition temperature ($T_{c}$), in the sense that the ratio $T_{c}/T_{F}$ is highest  among all known superfluids, where
 $T_{F}$ is the Fermi temperature. Since strong interactions are often the source of  macroscopic quantum effects, it is useful to have more examples of strongly  interacting systems.  By ``strongly interacting", we mean systems whose interaction energy is a significant part of the total energy at low temperatures. The interacting Fermi gases studied today are mostly associated with the so-called ``wide" resonances, whose widths are much larger than the Fermi energy $E_{F}$. In contrast,  there are very few studies on ``narrow" resonances\cite{Hulet}, for there is a  general belief that at all temperatures, the effect of a resonance could only be felt within its width $\Delta B$, and to stabilize the magnetic field within the width of a narrow resonance is a taunting task.

Here, we show that  such general view is incorrect. We shall see that in both high and low temperature regimes, the interaction energy associated with a narrow resonance is comparable to that of the unitary gas even when the system is many widths away from the resonance, as long as the distance from resonance $\gamma (B-B_{\infty})$ is within the relevant energy scale of the system ${\cal E}^{\ast}$, i.e. $\Delta B \ll \gamma (B-B_{\infty}) \ll {\cal E}^{\ast}$, where
${\cal E}^{\ast}$ is the Fermi energy $E_{F}$ at low temperatures, and temperature $T$ at high temperatures.
More specifically, we shall show that at low temperatures, the interaction energy is a substantial fraction of the total energy; 
and that at high temperatures,  the second virial coefficient $b_{2}$ at resonance is twice as large as that of a wide resonance, and remains sizable beyond its width. There is, however, a major difference between these two resonances. For wide resonances, the interaction energy of the scattering
states  ($\epsilon^{extended}$)  is antisymmetric about the resonance,
whereas it is highly asymmetric for narrow resonances -- strongly attractive on the atomic side, and weakly repulsive on the molecular side. Such differences are caused by a resonance structure of the phase shift unique to narrow resonances. The aforementioned large interaction energy and strong asymmetry should be observable experimentally.

{\em Wide and narrow resonance:} Currently, there are two schemes to
classify wide and narrow resonances. The first (denoted as ${\bf
(A)}$) is based on the property of the two-body systems\cite{Chin}.
The second (denoted as ${\bf (B)}$) makes use of the Fermi energy
$E_{F}$ of the many-body system
 \cite{Pethick,Petrov,HoDiener,Leo}.  They are not equivalent.
 To discuss their differences and relations, recall that the s-wave scattering length as
 a function of magnetic field is  $a_{s}(B) = a_{bg}\left( \frac{B-B_{0}}{B-B_{\infty}}\right)$, or
\begin{equation}
a_{s}(B) = a_{bg}\left( 1- \frac{\gamma \Delta
B}{\gamma(B-B_{\infty})}\right) \equiv a_{bg} - \frac{r_{\ast}
E_{\ast}}{\gamma (B-B_{\infty})} \label{asB}\end{equation} where
$a_{bg}$ is the background scattering length, $B_{0}$ and
$B_{\infty}$ are the magnetic fields at which $a_{s}$ vanishes and
diverges respectively;  $\Delta B\equiv B_{0} - B_{\infty}$  is the
width of the resonance, $\gamma$ is the magnetic moment difference
between the atom and closed molecular state. It can be shown from
two-channel models that $a_{bg}\gamma \Delta B
>0$\cite{Braaten}. We can then define a length $r_{\ast}$ as
$\hbar^2/(mr_{\ast} )\equiv a_{bg}\gamma \Delta B$, and energy scale $E^{\ast}=\hbar^2/mr^{\ast 2}$. The last equality in Eq.(\ref{asB}) is a more general representation as it applies to the case $a_{bg}=0$.  If $r_{\ast}$ is non-zero, then $\Delta B$ must diverge as  $a_{bg}$ vanishes.  The
region $B>B_{\infty}$ and $B<B_{\infty}$ will be referred to as the
``atomic" and ``molecular" side of the resonance, as a bound state
appears when $B<B_{\infty}$.

Eq.(\ref{asB}) shows that the resonance is characterized by the energies $E_{bg}=\hbar^2/(ma^{2}_{bg})$ and $\gamma\Delta B$. Their ratio will be denoted as
\begin{equation}
\alpha \equiv \frac{ \gamma \Delta B}{E_{bg}} = \frac{a_{bg}}{r_{\ast}}.
\end{equation}
In scheme {\bf (A)}\cite{Chin}, wide and narrow  resonance (denoted as $W_{\cal A}$ and $N_{\cal A}$) correspond to
\begin{equation}
{\cal W}_A : |\alpha| \gg 1,  \,\,\,\,\, {\cal N}_A :  |\alpha| \ll
1, \hspace{0.3in} {\rm Scheme} \, {\bf (A)}. \label{A}\end{equation}
In scheme {\bf (B)}\cite{Pethick,Petrov,HoDiener,Leo}, they
correspond to
\begin{equation}
{\cal W}_B :  r_{\ast}k_{F} \ll 1,  \,\,\,\,\, {\cal N}_B :
r_{\ast}k_{F}\gg  1,  \hspace{0.1in} {\rm Scheme} \, {\bf (B)}.
\label{B} \end{equation} Note that ${\bf (B)}$ makes no reference to
the width $\Delta B$, nor to the background scattering length
$a_{bg}$. In fact, many theoretical studies of ``narrow" resonances
make use a ``simple" two-channel models with $a_{bg}=0$, which
automatically sends  $|\Delta B|$  to $\infty$\cite{comment}. By
construction, such models can not produce the basic features of
$a_{s}(B)$ for all resonances that are experimentally regarded as
narrow, which all have very narrow width $\Delta B$. On the other
hand, as we shall see, such simple model does capture the essential
feature of the phase shift, which is what is needed to account for
the energy of the extended states, see footnote
\cite{comment-simple} later. Yet when it comes to many-body features
that depend crucially on $a_{bg}$ and $\Delta B$, it is necessary to go beyond
these simple models. To make contact with experiments, we shall
consider $a_{bg}\neq0$, with  $|a_{bg}|k_{F}<1$, i.e. the background
scattering is not near resonance.

\begin{figure}
\begin{center}
  \includegraphics[width=8.0cm,height=5.0cm]{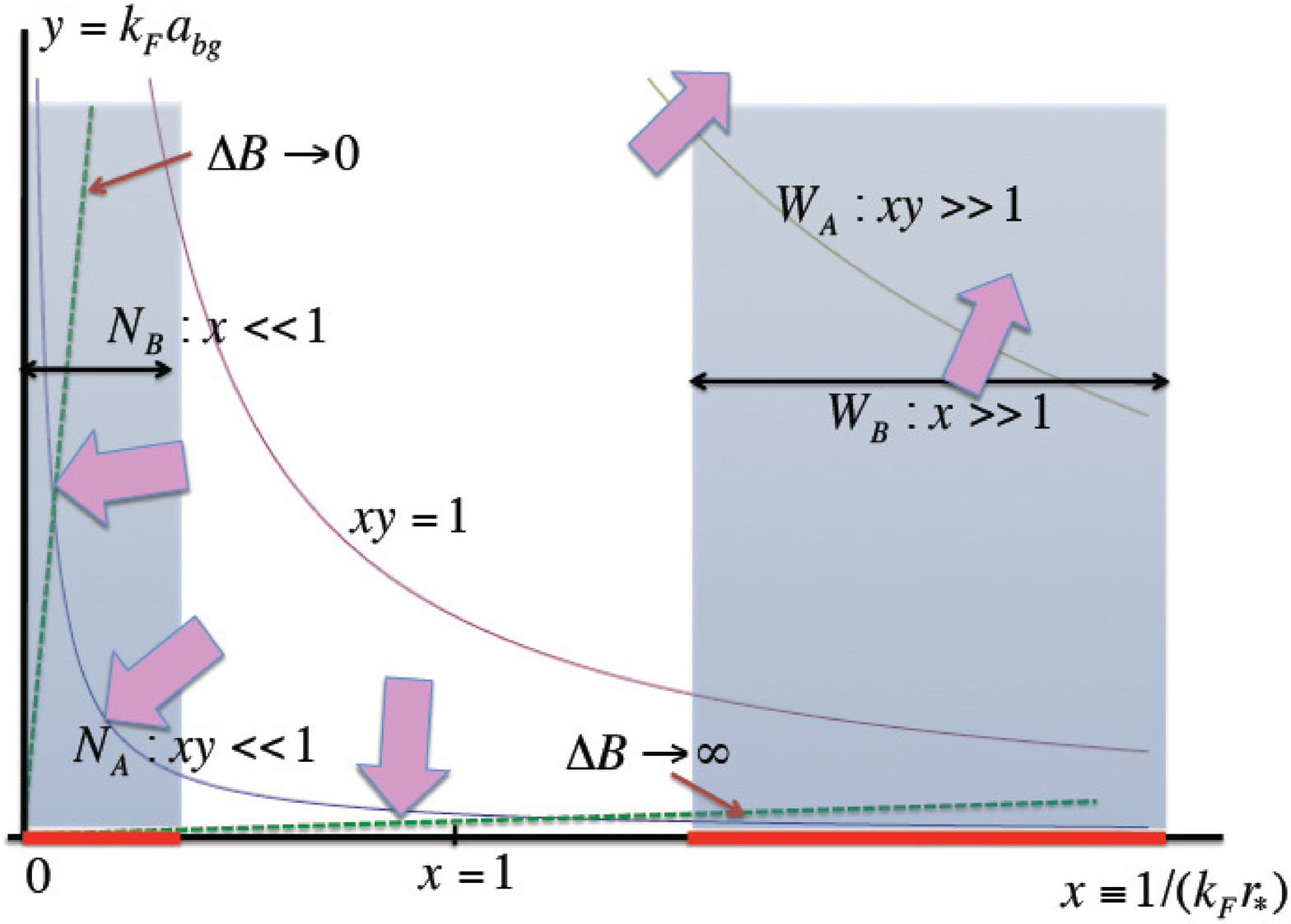}
    \caption{Wide and narrow resonances in the space $((k_{F}r_{\ast})^{-1}, k_{F}a_{bg})$:
 ${\cal W}_A$ and ${\cal N}_{A}$ are the regions above and below the curve $xy=1$ respectively.
 ${\cal W}_B$ and ${\cal N}_{B}$ are the regions where $x\gg1$ and $x\ll1$. The dotted line are contours of constant $\Delta B$. The thick red lines on the $x$-axis correspond to the simple two channel models\cite{comment}, (see text).  This phase diagram is for $a_{bg}>0$. The full phase diagram including  $a_{bg}<0$
 is mirror symmetric about the $x$-axis.
For $^{6}$Li  at density $n_{\uparrow}=n_{\downarrow}= 5\times 10^{14}$cm$^{-3}$,  its wide resonance at  834.1G has
 $a_{bg} =-1405a_{B}$,  $\Delta B=-300G$,  $|a_{bg}|/r_{\ast}=2.8\times 10^3$,   $r_{\ast} k_{F}= 8\times 10^{-4}$, and
  $k_{F}|a_{bg}|=2.3$. Its  narrow resonance at
 543.25G has
 $a_{bg} =61.6a_{B}$,   $\Delta B=0.1G$,  $a_{bg}/r_{\ast}=0.002$,  $r_{\ast}k_{F}=50$, and $k_{F}a_{bg}=0.1$\cite{Chin}.
}\label{phasediagram}
\end{center}
\end{figure}

The relation between scheme ${\bf (A)}$ and ${\bf (B)}$ in the space of $( r_{\ast}^{-1}, a_{bg})$ is shown in Figure 1. Defining the dimensionless quantities $x = (k_F r_{\ast})^{-1}$ and $y=k_{F} a_{bg}$,
scheme ${\bf (A)}$ and ${\bf (B)}$ correspond to the follow regions in the (x,y) plane, (${\cal W}_A$:  $xy=\alpha \gg 1$ ; ${\cal N}_A$:  $xy\ll 1$) and (${\cal W}_B$:  $x\gg 1$ ; ${\cal N}_B$:  $x\ll 1$). (See Figure 1). The contours of constant $\Delta B$ are denoted by dotted straight lines, whose slope $y/x$ is
the ratio $2E_F/(\gamma\Delta B)$.
 The regions where the simple two channel model (with $a_{bg}=0$ and hence $\Delta B=\infty$) applies are along the $x$-axis, indicated by thick red lines.

 {\em Origin of Strong Interaction:}
The effects of interaction can be found in the two body phase shift  $\delta(k)$.
From the general two channel model which includes the background scattering length $a_{bg}$, one can calculate the scattering amplitude  $f(k)= 1/(k {\rm cot}\delta(k) - ik)$ and finds that the energy dependent scattering length $a(k)$, defined as $ k {\rm cot}\delta(k) \equiv -1/a(k)$, has the general form\cite{Braaten}
\begin{equation}
  a(k) = a_{bg}\left( \frac{\hbar^2k^2/m  -  \gamma (B-B_{0})}{\hbar^2k^2/m  - \gamma (B-B_{\infty})}\right),
 \label{ak}  \end{equation}
 and $a(k=0)$ is the s-wave scattering length whose magnetic field dependence is given in Eq.(\ref{asB}).
 It is straightforward to show from
Eq.(\ref{ak}) that
\begin{equation}
{\rm tan}\delta(k) = -ka_{bg} - \frac{ k \hbar^2/mr_{\ast}}{ \hbar^2
k^2/m - \gamma (B-B_{\infty})}. \label{tand}
\label{tandelta} \end{equation}
Since $\delta(k)$ is defined up to modulo $\pi$, its magnetic field
dependence is to be determined by Levinson's theorem -- that it is a
continuous function of $B$ until a bound state emerges, in which
case it jumps down by $-\pi$.   Applying the conditions in Eq.(\ref{A}) and (\ref{B})
to Eq.(\ref{tand}), we have plotted  $\delta(k)$ for wide and narrow
resonance   for various magnetic fields  in Figure 2A and 2B
respectively.

\begin{figure}
  \begin{center}
  \includegraphics[width=7.0cm,height=5.0cm]{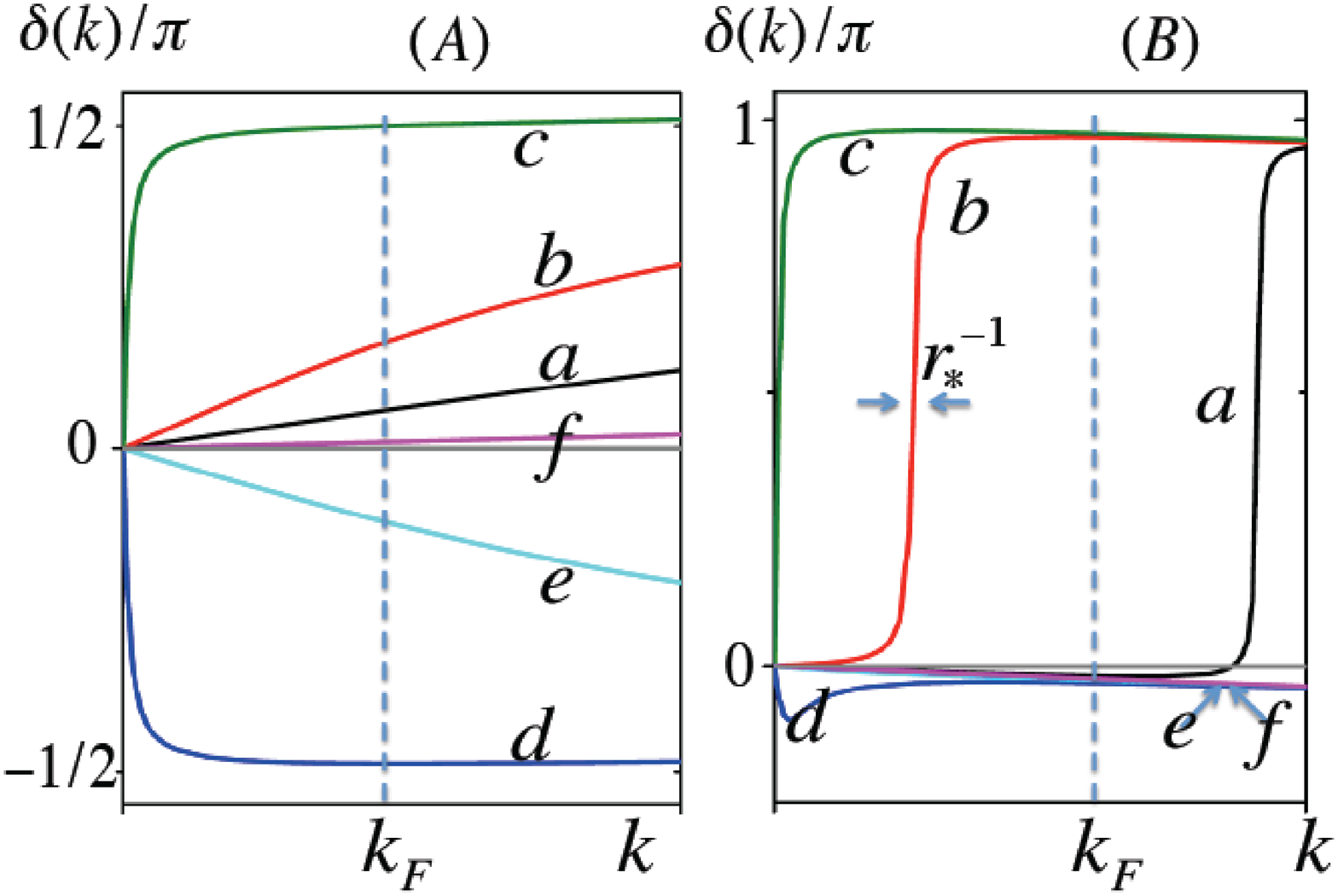}
    \caption{$\delta(k)$ vs $k$ for wide and narrow resonances: The labels  (a,b,c,d,e,f) correspond to ($a : B\gg B_{\infty}$; $b : B> B_{\infty}$; $c : B= B_{\infty}+0^{+}$; $d : B= B_{\infty}-0^{+}$; $e : B < B_{\infty}$; $f : B\ll B_{\infty}$). For wide resonance with $a_{bg}<0$, (Fig. 2A), we have $\delta(k) = -{\rm arctan}(ka_{s}(B))$. Near resonance, (c and d), $\delta(k)$ approaches a step function of height $\pm\pi/2$. Far from resonance, (a and f) $\delta(k)$ reduces to $\delta(k) = -{\rm arctan}(ka_{bg})\sim - k a_{bg}$, and 
    $|k a_{bg}|\ll 1$.
For narrow resonance with $a_{bg}>0$,  (Fig. 2B), when $B>B_{\infty}$, $\delta(k)$ approaches a step function of height $\pi$ with a width $1/r_{\ast}$. For $B<B_{\infty}$, $\delta(k)$ quickly reduces to $-ka_{bg}$ as $B_{\infty}-B$ exceeds $\Delta B$, and $| ka_{bg}|\ll 1$. 
The $\delta(k)$ for both $e$ and $f$ are essentially identical.
}\label{delta}
 \end{center}
\end{figure}

Figure 2A shows that as one approaches a wide resonance from the
atomic side, $\delta (k)$ starts from the form $-a_{bg}k$
(which is much less than 1 for $k$ up to a few $k_{F}$);
and turns into a step function of height $\pi/2$. In the latter case, 
{\em all} energy states to gain a $\pi/2$ phase shift,  leading to a
large negative interaction energy. Crossing the resonance to the molecular side, $\delta(k)$ 
jumps down by $-\pi$ due to the appearance of a bound state, leading to a
repulsive energy on the molecular side.


For a narrow resonance,  $\delta (k)$ is essentially a step function of height $\pi$ located at energy $\hbar^2 k^2/m= \gamma (B-B_{\infty})$ with a width $\Delta k = 1/r_{\ast}$, as shown in Figure 2B.
We can see that even when  $\gamma \Delta B\ll \gamma (B-B_{\infty})\ll 2E_{F}=\hbar^2 k_{F}^2/m$ or $
T$, i.e.  the system is far beyond the width of the resonance, all states with energy above $\gamma (B-B_{\infty})$ are phase shifted by  $\pi$, (twice the value of wide resonance); thereby generating
considerable interaction energy.
As $B$ passes through to the molecular side, $\delta (k)$ shifts $\pi$ down to zero due to the
the presence of a bound state, leading to a small interaction energy for the scattering state.

\begin{figure}
    \includegraphics[width=8.0cm, height=5.0cm]{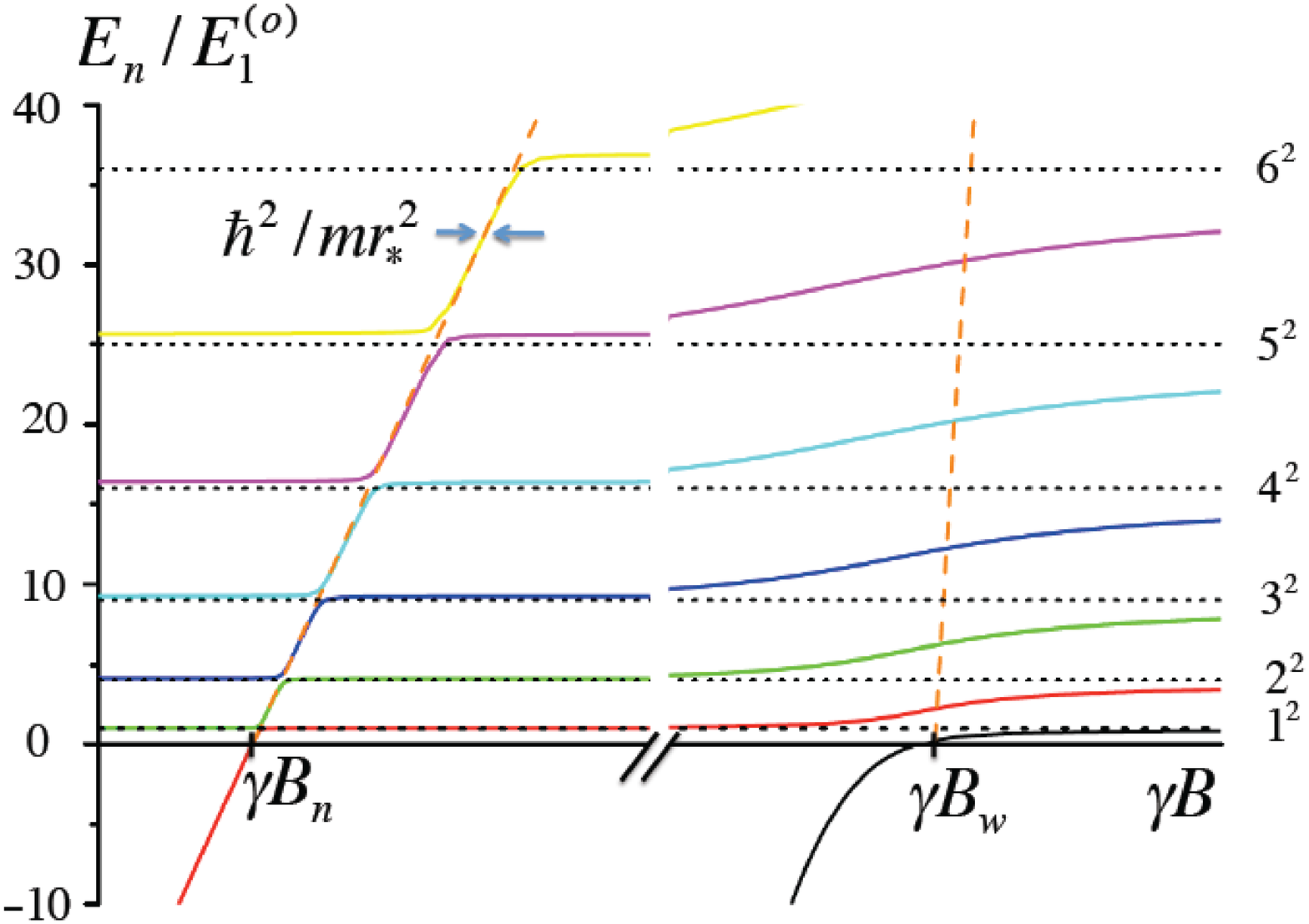}
    \caption{The evolution of energy level as $B$ passes through a wide resonance (at $B_w$) and
    a narrow resonance (at $B_{n}$):
For wide resonance, the transition from $k_{n}^{(o)}$ to $k_{n-1}^{(o)}$ is completed only half way at resonance. 
 For narrow resonance, due to  the narrow width, this transition is completed on the atomic side of the resonance, before the bound state is formed.  Passing the resonance to the molecular side, ($B<B_{\infty}$), $\delta (k)<<1$ and the scattering state has little interaction energy. }\label{fig4}
\end{figure}

A more detailed understanding of the origin of interaction energy can be obtained from the evolution of the energy levels as they pass through a resonance. Fig.\ref{fig4} shows such an evolution for the passage of a wide followed by that of a narrow resonance, as we note that
many atoms have narrow resonances close to wide ones.  The energy levels $\hbar^2 k^2/m$ are calculated through standard methods by setting the
wavefunction of the extended state $\psi(r) = r^{-1}{\rm sin}(kr + \delta(k))$ to zero at a large distance $r=R$, which gives
$ k_{n} = k_{n}^{(o)} - \delta(k_{n})/R$, where $k_{n}^{(o)}= n\pi/R$, $n=1,2, ...$ are the momenta of non-interacting systems.

As one moves from right to left in Fig.3, the energy levels shift down from $k_{n}^{(o)}$ to $k_{n-1}^{(o)}$, with the lowest level developed into a bound state.  
For the wide resonance,  
the transition from $k_{n}^{(o)}$ to $k_{n-1}^{(o)}$  ($n\geq 1$) is only completed half way at resonance, due to the fact that $\delta = \pi/2$ at resonance. In contrast, for a narrow resonance located at $B_{n}$, $k_{\ell}^{(o)}$  shifts down to $k_{\ell-1}^{(o)}$ rapidly (due to the narrow width) on the atomic side of the resonance 
($B>B_{n}$) at $\gamma (B-B_{n}) = \hbar^2 k_{\ell}^{(o) 2}/m$.  As a result, all momentum states with energies above $\gamma (B-B_{n})$ are phase shifted by $\pi$, thereby generating considerable interaction energy. On the other hand, on the molecular side of the resonance ($B<B_{n}$), the phase shift becomes very small (see curves $d$, $e$, $f$ in figure 2B). The energy levels therefore becomes very close to the non-interacting energy level as shown in Fig.3, implying little interaction energy on the molecular side.

Recently, it has been shown in ref.\cite{KK} that the energy dependence of the phase shift of ``narrow" resonances can lead to a high  $T_{c}/T_F$ ratio. The resonance they studied, however, has $r_{\ast}k_{F}<1$. It is therefore not narrow in the present  as well as previous classifications\cite{Pethick,Petrov,HoDiener,Leo}. The origin of strong interaction, and the implication of the  $\pi$-jump in the phase shift of narrow resonances were not discussed in ref.\cite{KK}. 

\begin{figure}
  \begin{center}
    \includegraphics[width=7.5cm, height=5.0cm]{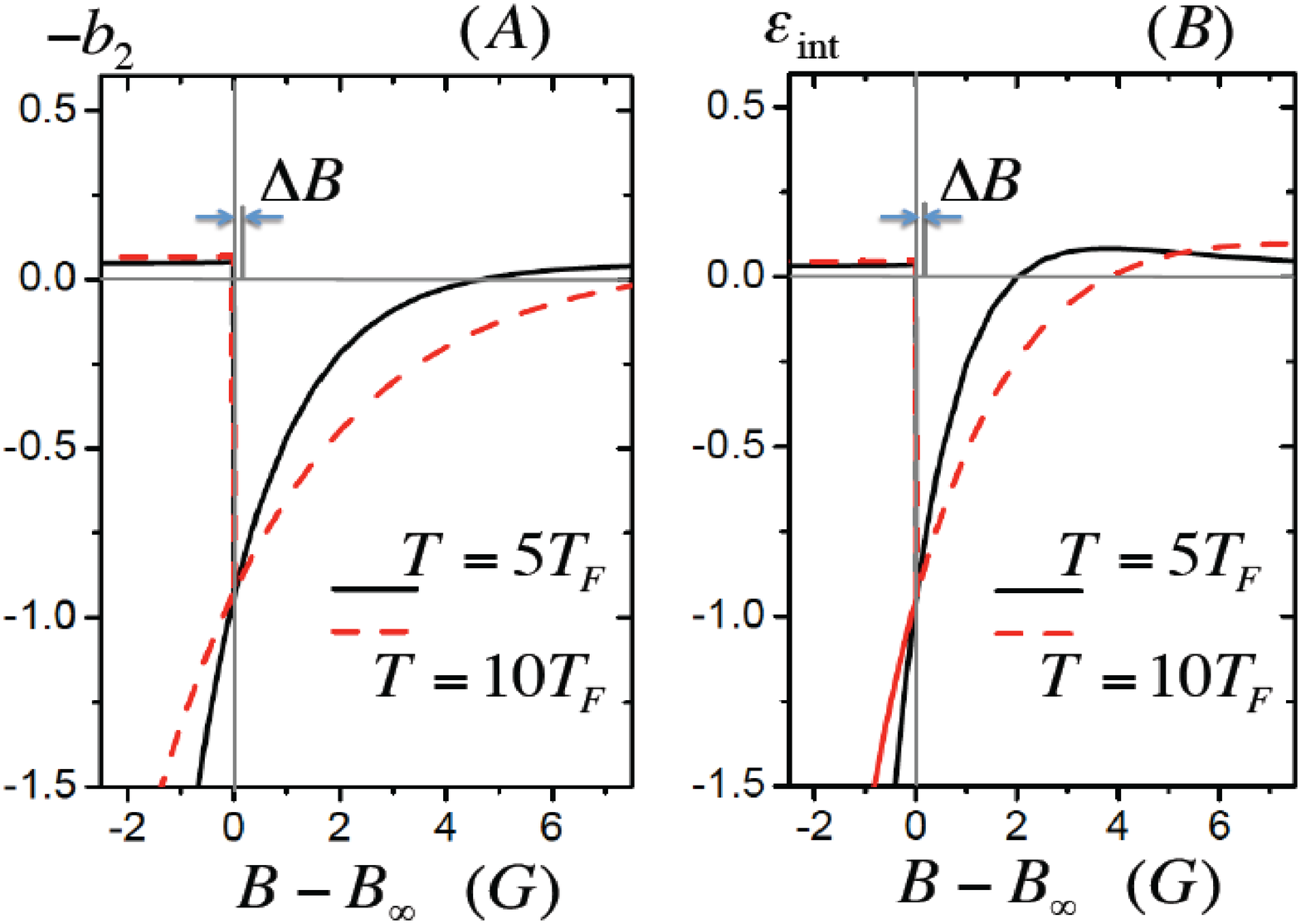}
    \caption{$-b_2$(A) and $\epsilon_{int}$(B) as a function of magnetic field:  We consider
    a gas of  $^{6}$Li at $543.25G$ with parameters given the caption of Figure 1.
In this case, $T_{F}=40\mu K \sim 3\gamma \Delta B$ (in temperature units). Note that  $b_{2}$ remains sizable even when 
    $B-B_{\infty} \sim 2G \sim 20\Delta B$. This   is because we are at $T=5T_{F}\sim 15 \gamma \Delta B$ or larger, where thermal sampling extends over an energy range of several $T$.
    When the system is in equilibrium, it follows the lower curve (or the ``lower branch"). At resonance and in the limit $\Delta B\rightarrow 0$,
$b_2\rightarrow 1$. On the molecular side, $B<B_{\infty}$,
$\epsilon_{int}^{extended}$ (for the upper branch) is given by the flat line, given by the small value $\delta \sim - k a_{bg}$. }\label{b2}
 \end{center}
\end{figure}

{\em Virial expansion:} A useful gauge of the interaction effects of narrow resonances is to compare them with those of wide resonances. At high temperatures, the interaction energy of a gas with density $n$ is given by \begin{equation}
\epsilon_{int}(T,n) = \frac{3Tn}{2}\left(\frac{n\lambda^3}{\sqrt{2}}\right)
\left[ -b_{2} + \frac{2T}{3}
\frac{\partial b_{2}}{\partial T} \right],
\end{equation}
where $b_{2}$ is the second virial coefficient. It consists of two contributions,
 $b_{2} = b_{2}^{bound} + b_{2}^{extended}$, where $b_{2}^{bound}=  \sum_{\alpha}e^{|E^{b}_{\alpha}|/T}$ is the partition function of the bound states $\{ E_{\alpha}^{b}\}$ (labelled by $\alpha$) and
\begin{equation}
b_{2}^{extended} = \int^{\infty}_{0} \frac{{\rm d}k}{\pi} \frac{{\rm d}\delta(k)}{{\rm d} k} e^{-\hbar^2 k^2/mT}\label{b2extended}
\label{b2exteneded} \end{equation}
is the change of the partition function of the scattering states. Their corresponding contributions to interaction energy will be denoted as $\epsilon_{int}= \epsilon_{int}^{bound} +  \epsilon_{int}^{extended}$ respectively.  For wide resonances, $b_2$ and $\epsilon_{int}$ have been studied in ref.\cite{HoMueller}. In Figure 4A and 4B, we have plotted both $-b_{2}$ and $\epsilon_{int}$ for narrow resonance. They are the curves that bend downward. The flat lines that exist only on the molecular side of the resonance are
$b_{2}^{extended}$ and $\epsilon^{extended}_{int}$. They contain only the contributions from the scattering state and are sometimes referred to as  ``upper" or ``repulsive" branch quantities. 
The full $b_2$ and $\epsilon_{int}$ consisting of both scattering state and bound state contributions are referred to as ``lower" or ``attractive" branch quantities. 
Note that the interaction energy of the repulsive branch is close to 0, $\epsilon^{extended}_{int}\sim 0$, whereas that of the attractive branch near resonance is  large and negative.
This is due to the special property of $\delta(k)$ discussed in the previous section.
Comparing the interaction energy of narrow resonance (in Fig.4) with that of wide resonance\cite{HoMueller}, we note that the range of $B$ field over which $b_2$ is non-zero is much smaller than that of the wide resonance. Despite that, the large value of $b_2$ still  extends far beyond the width of the resonance, and its value at resonance is 1, twice of that of a wide resonance. This is a consequence of the $\pi$ jump of $\delta(k)$ at energy $\hbar^2 k^2/m=\gamma (B-B_{\infty})$ shown in Fig.2B.

{\em Interaction energy at low temperatures:}  To extend the study of upper and lower branch energy to lower temperatures, we apply a generalization of Nozieres-Schmitt-Rink (NSR) method to calculate the energy of the system\cite{ShenoyHo}.  Central to the NSR method is the $T$-matrix ($T({\bf q}, \omega)$) in a medium for a pair of fermions with total momentum ${\bf q}$.  In the case of narrow resonance, it is
\begin{equation}
\frac{1}{T({\bf q},\omega)}= \frac{m}{4\pi a_{s}(k)} - \frac{1}{\Omega}\sum_{\bf p}\left(  \frac{\gamma({\bf p, q}) }{\omega_{+} - \omega(q)-\frac{p^2}{m}} -\frac{1}{\frac{p^2}{m}}\right),
\end{equation}
where $k^2/m=\omega- \omega(q)$, $\omega(q)= q^2/(4m)-2\mu$; $\omega_{+}= \omega + i 0^{+}$, $\gamma({\bf p, q})= 1-n_{F}(\xi_{{\bf q}/2+{\bf p}}) - n_{F}(\xi_{{\bf q}/2-{\bf p}})$ describes the Pauli blocking effect of the medium, $\xi_{{\bf p}} = p^2/(2m) -\mu$ and we have set $\hbar=1$. From the phase of inverse $T$-matrix, $\delta({\bf q}, \omega) = {\rm arg}T^{-1}({\bf q}, \omega)$, one can calculate the pressure, the energy density, and equation of state of the attractive  branch and the repulsive branch. 
In Fig.5, we have plotted the result for the interaction energy for both the attractive and repulsive branch  across resonance at a temperature $T=0.5T_{F}$. The behaviors of both branches are similar to those found from the virial expansion (Fig. 3). The energy scale, however, is very different. One sees that the interaction energy is as much as 50$\%$ of the total energy  of an ideal Fermi gas right at resonance, and can be as high as $30-40\%$ {\em even beyond the width of the resonance.}
In Fig. 5, one also see that the interaction energy is only significant when
the distance from resonance $\gamma(B-B_{\infty})$ is within $2E_{F}$, as discussed previously.

\begin{figure}
  \begin{center}
    \includegraphics[width=7.5cm, height=5.5cm]{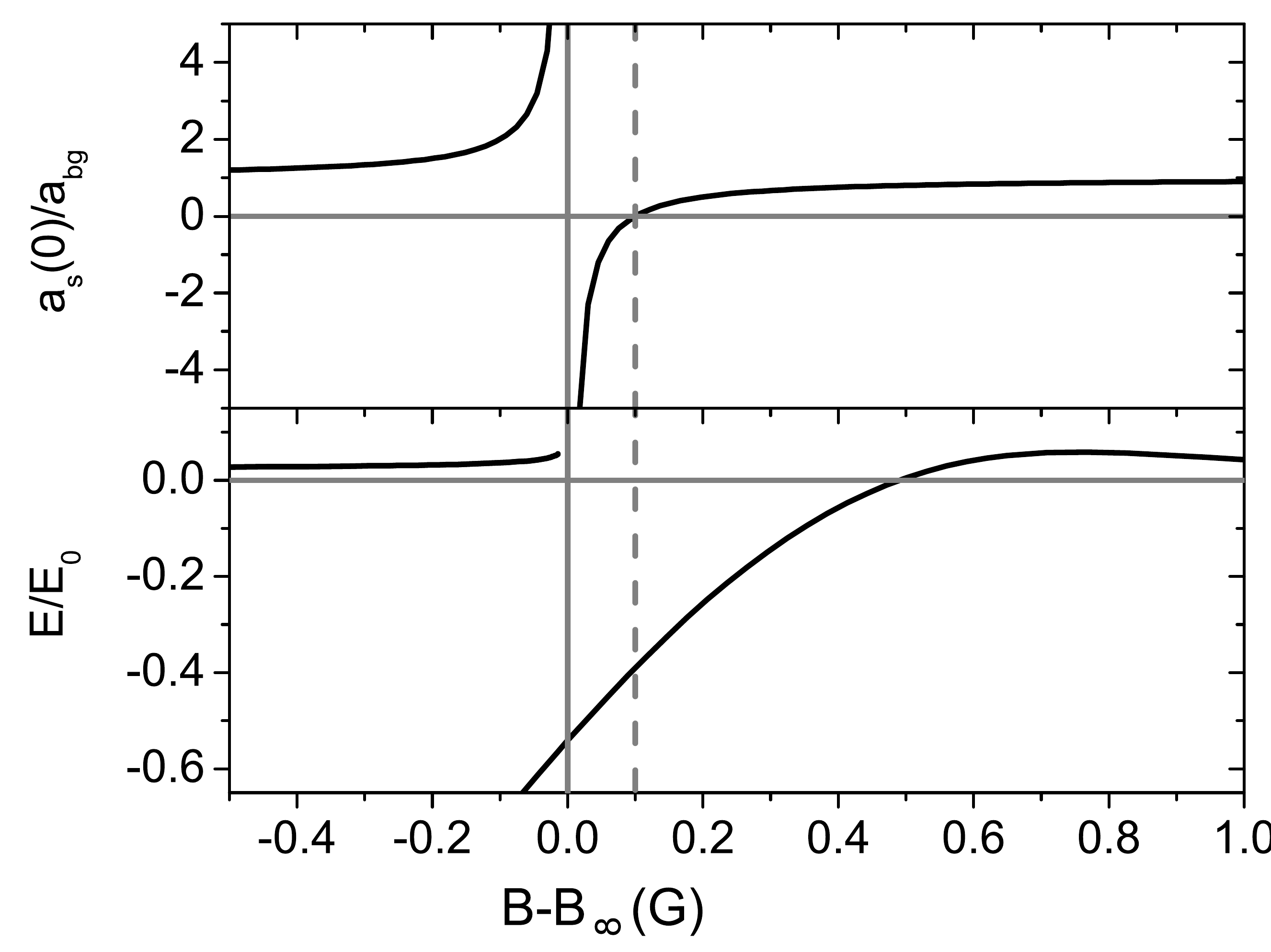}
      \caption{S-wave scattering length(upper panel) and interaction energy (lower panel) of $^{6}$Li across $543.25G$ narrow resonance at low temperatures. We have $T=0.5T_F$,  $T_F=3\gamma \Delta B$ as in Fig.4. Other parameters are the same as those given in the caption of Fig.1. $E_0$ is the energy for a non-interacting Fermi gas at the same temperature. The dashed line indicates the width $\Delta B$ of the resonance.  The downward turning curve and the flat curve are the energies of the lower and upper branch respectively.  The interaction energy of the lower branch reaches 50$\%$ of the free fermion energy at resonance, and remains sizable beyond the width of the resonance.}\label{b2}
 \end{center}
\end{figure}


After the posting of an earlier version of this paper which does not include the low temperature results, Ken O'hara's group has performed rf spectroscopy studies on the narrow resonance of $^{6}$Li at $543.25$G and has found the asymmetry of interaction energy $\epsilon_{int}$ on different side of the resonance\cite{Ohara}. 
We note that $\epsilon_{int}$ can also be determined exactly (free of the modeling by specific theories) from in situ density measurements\cite{HoZhou}.

We thank Vijay Shenoy for stimulating discussions. This work is supported by NSF Grant DMR-0907366 and by DARPA under
the Army Research Office Grant Nos. W911NF-07-1-0464, W911NF0710576, and the Tsinghua University Initiative Scientific Research Program.


\begin{thebibliography}{99}
\bibitem{comment}This is the hamiltonian $H=\sum_{\bf k, \sigma}\epsilon_{\bf k}c^{\dagger}_{{\bf k}, \sigma}c_{{\bf k}, \sigma}
+\nu d^{\dagger}d
+\alpha\sum_{\bf k} (d^{\dagger}c^{}_{{\bf k}, \uparrow}c^{}_{{\bf k}, \downarrow} + h.c.) $, where $\nu$ is the bare detuning, and $\alpha$ is the coupling between the close channel molecule $d$ and the open channel fermions $c_{{\bf k}\sigma}$. This model implies $4\pi \hbar^2 a_{s}/m =-|\alpha|^2/\overline{\nu}$ with $a_{bg}=0$, where
$\overline{\nu}$ is the physical detuning.
The many-body property of this narrow resonance model is discussed in ref.\cite{Leo}.
\bibitem{comment-simple}Eq.(\ref{tandelta}) in the limit of $a_{bg}\rightarrow 0$ with $r_{\ast}$ held fixed (hence $\Delta B\rightarrow \infty$) yields the same form as that calculated from the simple model in \cite{comment}.
\end{thebibliography}
\end{document}